\documentclass[11pt,fleqn]{article}
\usepackage{amsfonts, amsmath, epsfig, color, verbatim}
\usepackage[final]{changes}
\newcommand{\ol}{\setlength{\itemsep}{0pt.}\begin{enumerate}}
\newcommand{\eol}{\end{enumerate}\setlength{\itemsep}{-\parsep}}
\newcommand{\ignore}[1]{}
\title{On Coset Leader Graphs of LDPC Codes}
\author{Eran Iceland and Alex Samorodnitsky
\thanks{School of Engineering and Computer Science,
The Hebrew University of Jerusalem,
Jerusalem 91904, Israel.
Research partially supported by ISF
grant 1241/11 and by BSF grant 2010451. }
}

\begin{document}
\maketitle


\newtheorem{THEOREM}{Theorem}[section]
\newenvironment{theorem}{\begin{THEOREM} \hspace{-.85em} {\bf :}
}%
                        {\end{THEOREM}}
\newenvironment{thm1}{\begin{THEOREM} \rm}
                     {\end{THEOREM}}
                        
\newtheorem{LEMMA}[THEOREM]{Lemma}
\newenvironment{lemma}{\begin{LEMMA} \hspace{-.85em} {\bf :} }%
                      {\end{LEMMA}}
\newtheorem{COROLLARY}[THEOREM]{Corollary}
\newenvironment{corollary}{\begin{COROLLARY} \hspace{-.85em} {\bf
:} }%
                          {\end{COROLLARY}}
\newtheorem{PROPOSITION}[THEOREM]{Proposition}
\newenvironment{proposition}{\begin{PROPOSITION} \hspace{-.85em}
{\bf :} }%
                            {\end{PROPOSITION}}
\newtheorem{DEFINITION}[THEOREM]{Definition}
\newenvironment{definition}{\begin{DEFINITION} \hspace{-.85em} {\bf
:} \rm}%
                            {\end{DEFINITION}}
\newtheorem{EXAMPLE}[THEOREM]{Example}
\newenvironment{example}{\begin{EXAMPLE} \hspace{-.85em} {\bf :}
\rm}%
                            {\end{EXAMPLE}}
\newtheorem{CONJECTURE}[THEOREM]{Conjecture}
\newenvironment{conjecture}{\begin{CONJECTURE} \hspace{-.85em}
{\bf :} \rm}%
                            {\end{CONJECTURE}}
\newtheorem{MAINCONJECTURE}[THEOREM]{Main Conjecture}
\newenvironment{mainconjecture}{\begin{MAINCONJECTURE} \hspace{-.85em}
{\bf :} \rm}%
                            {\end{MAINCONJECTURE}}
\newtheorem{PROBLEM}[THEOREM]{Problem}
\newenvironment{problem}{\begin{PROBLEM} \hspace{-.85em} {\bf :}
\rm}%
                            {\end{PROBLEM}}
\newtheorem{QUESTION}[THEOREM]{Question}
\newenvironment{question}{\begin{QUESTION} \hspace{-.85em} {\bf :}
\rm}%
                            {\end{QUESTION}}
\newtheorem{REMARK}[THEOREM]{Remark}
\newenvironment{remark}{\begin{REMARK} \hspace{-.85em} {\bf :}
\rm}%
                            {\end{REMARK}}

\newcommand{\thm}{\begin{theorem}}
\newcommand{\lem}{\begin{lemma}}
\newcommand{\pro}{\begin{proposition}}
\newcommand{\dfn}{\begin{definition}}
\newcommand{\rem}{\begin{remark}}
\newcommand{\xam}{\begin{example}}
\newcommand{\cnj}{\begin{conjecture}}
\newcommand{\mcnj}{\begin{mainconjecture}}
\newcommand{\prb}{\begin{problem}}
\newcommand{\que}{\begin{question}}
\newcommand{\cor}{\begin{corollary}}
\newcommand{\prf}{\noindent{\bf Proof:} }
\newcommand{\ethm}{\end{theorem}}
\newcommand{\elem}{\end{lemma}}
\newcommand{\epro}{\end{proposition}}
\newcommand{\edfn}{\bbox\end{definition}}
\newcommand{\erem}{\bbox\end{remark}}
\newcommand{\exam}{\bbox\end{example}}
\newcommand{\ecnj}{\bbox\end{conjecture}}
\newcommand{\emcnj}{\bbox\end{mainconjecture}}
\newcommand{\eprb}{\bbox\end{problem}}
\newcommand{\eque}{\bbox\end{question}}
\newcommand{\ecor}{\end{corollary}}
\newcommand{\eprf}{\bbox}
\newcommand{\beqn}{\begin{equation}}
\newcommand{\eeqn}{\end{equation}}
\newcommand{\wbox}{\mbox{$\sqcap$\llap{$\sqcup$}}}
\newcommand{\bbox}{\vrule height7pt width4pt depth1pt}
\newcommand{\qed}{\bbox}
\def\sup{^}

\def\H{\{-1,1\}^n}
\def\B{\{0,1\}^n}

\def\S{S(n,w)}

\def\n{\lfloor \frac n2 \rfloor}

\def \E{\mathbb E}
\def \R{\mathbb R}
\def \N{\mathbb N}
\def \F{\mathbb F}
\def \T{\mathbb T}

\def\<{\left<}
\def\>{\right>}
\def \({\left(}
\def \){\right)}
\def \e{\epsilon}
\def \r{\rfloor}

\def \1{{\bf 1}}

\def \noi{\noindent}

\newcommand{\rarrow}{\rightarrow}

\newcommand{\larrow}{\leftarrow}

\overfullrule=0pt
\def\setof#1{\lbrace #1 \rbrace}

\begin{abstract}
Our main technical result is that\replaced{,}{ metric balls} in the coset leader graph of a linear binary code of block length $n$\added{, the metric balls} spanned by constant-weight vectors grow exponentially slower than those in $\B$.

Following the approach of \cite{friedman2006generalized}, we use this fact to improve on the first linear programming bound on the rate of LDPC codes, as the function of their minimal \added{relative }distance. This improvement, combined with the techniques of \cite{ben2006upper}, improves the rate vs distance bounds for LDPC codes in a significant sub-range of relative distances.
\end{abstract}

\section{Introduction}

\noi This paper deals with rate versus distance bounds for binary error-correcting codes.

\noi A binary code $C$ of block length $n$, rate $R$, and relative minimal distance $\delta$ is a subset of $\B$ of cardinality $2^{Rn}$, such that the Hamming distance between any two distinct elements of $C$ is at least $d = \delta n$. A fundamental open problem in coding theory is to find the largest possible asymptotic rate $R = R(\delta)$ for which there exists a family of codes $\left\{C_n\right\}_n$ with block length $n \rarrow \infty$, rate at least $R$ and \added{relative }distance at least $\delta$.

\noi The best known bounds on $R(\delta)$ are
\[
1 - H(\delta) \le R(\delta) \le R_{LP}(\delta)
\]
The first inequality is the Gilbert-Varshamov bound \cite{macwilliams1977theory}. Here $H(\cdot)$ is the binary entropy function. In the second inequality, we denote by $R_{LP}(\delta)$ the {\it second JPL bound} \cite{mceliece1977new}, obtained via the linear programming approach of Delsarte \cite{delsarte1973algebraic}. For an explicit expression for $R_{LP}(\delta)$ see e.g., \replaced{\cite{macwilliams1977theory}} {\cite{levenshtein1998universal}}.

\noi Linear codes are an important subclass of error-correcting codes. A linear code of rate $R$ is an $Rn$-dimensional linear subspace of $\B \cong \F^n_2$.

\noi In this paper we consider a special class of linear codes. These are the \replaced{Low-Density Parity Check (LDPC)}{LDPC (Low-Density Parity Check)} codes. An LDPC code $C$ comes with an additional parameter - an absolute constant $w$. It has an additional structure: the dual code (dual subspace) $C^{\perp}$ is spanned by vectors of weight at most $w$.

\noi LDPC codes were introduced by Gallager \cite{gallager1963low}. They are important both in theory and in practice of robust communications. A question of interest is to investigate the rate vs. minimal distance dependence in this class of codes. Let $R_w(\delta)$ be the largest possible asymptotic rate of an LDPC code whose dual is spanned by vectors of weight $w$ or less.

\noi Gallager has shown that, for large $w$, LDPC codes reach the Gilbert-Varshamov bound, that is
\[
\limsup_{w \rarrow \infty} R_w(\delta) \ge 1 - H(\delta)
\]
From the other side, upper bounds on $R_w(\delta)$ were obtained in \cite{burshtein2002upper, ben2006upper}. These papers use the linear programming framework, combined with direct combinatorial and information-theoretic arguments exploiting the special structure of $C^\perp$, to improve on the second JPL bound $R_{LP}(\delta)$ for all values of $\delta$.

\noi This paper continues the line of research started in \cite{burshtein2002upper, ben2006upper}. Our starting point is the elegant proof of the {\it first JPL bound}\footnote{This bound, also proved in \cite{mceliece1977new}, coincides with the second JPL bound for $0.273... \leq \delta \leq 1/2$.} \added{$ R(\delta) \le H(1/2 - \sqrt{\delta(1-\delta)})$} for linear codes given in \cite{friedman2006generalized}. Given a linear code $C$, the strategy is to compare metric spaces defined on two graphs: the discrete cube $\B$ and the {\it coset leader} graph \replaced{$\T$ defined as the Cayley graph of the quotient group $\F_2^n / C^\perp$ with respect to the set of generators given by the standard basis $e_1 \ldots e_n$.} {$\T = \B/C^\perp$. Recall that we impose a multi-graph structure on $\T$ by connecting a coset $x + C^{\perp}$ in $\T$ to the elements of the multiset $\left\{x + e_j + C^{\perp}\right\}$, $j = 1,...,n $.} \added{If $e_i + e_j \in C^\perp$, then edges in directions $i$ and $j$ are parallel and $\T$ becomes a multi-graph.}\footnote{\added{In what follows, we treat both cases exactly in the same way. Hence it might be easier for the reader always to think of $\T$ as a simple graph.}}

\noi \added {The name 'coset leader graph' comes from a well-known notion in coding theory. Recall that a minimal weight element in a coset is called the {\it coset leader} \cite{macwilliams1977theory}. (If the coset has more than one element of minimal weight, we take the coset leader to be minimal in the lexicographic order among them). This establishes a one-to-one correspondence between the vertices of $\T$ and coset leaders of $C^\perp$.}

\noi \replaced{For a graph $G$, a vertex $x\in G$, and an integer parameter $r$, the metric ball $B(x,r)$ is the set of vertices whose distance from $x$ in the graph metric is at most $r$. We will be interested in the rate of growth of metric balls in $\T$. Since $\T$ is a vertex-transitive graph, we may choose the center arbitrarily, and we fix it to be the coset of zero. Accordingly, let $B_{\T}(r)$ be the metric ball $\left\{x\in \T~:d(x,C^\perp) \le r\right\}$. (Note that $B_{\T}(r)$ is the set of cosets with coset leader of Hamming weight at most $r$.) We are motivated by the following result of \cite{friedman2006generalized} restated in our own words.}{In particular, we will be interested in how fast the metric balls $B_\T(r) = \left\{x\in \T~:d(x,0) \le r\right\}$ in $\T$ grow as a function of their radius $r$, compared to the corresponding growth in $\B$. The motivation is the following (streamlined version of) result of \cite{friedman2006generalized}.}

\begin{thm1} { (\cite{friedman2006generalized}){\bf:}}
\label{thm:FT}
Let $C$ be a linear code with relative minimal distance $\delta$. Let $\T = \B/C^\perp$ be the coset leader graph of $C^{\perp}$. Set $r = \(\frac12 - \sqrt{\delta(1-\delta)}\) \cdot n$. Then
\[
|C| \le 2^{o(n)} \cdot  | B_\T(r)|
\]
\end{thm1}

\noi Our main technical result is that if $\T$ comes from an LDPC code, then the growth \added{of metric balls} in $\T$ is exponentially slower \added {than that in $\B$. Let $B(r)$ be the Hamming ball of radius $r$ in $\B$ centered at zero. That is, $B(r) = \left\{x\in \B~: |x| \le r\right\}$, where $|\cdot|$ denotes the Hamming weight.}
\thm
\label{thm:tor_size}
For any integer \replaced{$w \ge 3$} {$3 \le w$} and $0 < \rho < 1/2$,\footnote{The case $w=2$ is not interesting since it is easy to see that $R_2(\delta) = 0$ for any $\delta >0$.} there is a constant $c = c(w,\rho) \replaced{\geq \frac{\log_2e}{8w^2} \cdot \(\frac{\rho^w}{2}\)^{w+1}}{>0}$ such that the following holds for any $n \ge w$:

\noi Let $C \subseteq \B$ be a linear code whose dual code $C^{\perp}$ is spanned by vectors of \replaced{weight}{length} at most $w$, and let $\T = \B/C^{\perp}$.

\noi Then
\begin{equation} \label{eq:tor_size}
|B_{\T}(\rho n )| \le 2^{-cn} \cdot |B(\rho n)|.
\end{equation}
\ethm

\noi Taken together with Theorem~\ref{thm:FT}, this implies our main result\deleted{, an improved upper bound on $R_w(\delta)$}. \added{Recall that the first JPL bound is $ R(\delta) \le H\(1/2 - \sqrt{\delta(1-\delta)}\)$. We improve this bound for $R_w(\delta)$.}

\cor
\label{cor:LP-for-LDPC}
For any $w \geq 3$,
\[
R_w(\delta) \le H\(\frac12 - \sqrt{\delta(1-\delta)}\) - c\(w,~\frac12 - \sqrt{\delta(1-\delta)}\)
\]
where $c(w,\rho)$ is given by Theorem~\ref{thm:tor_size}.
\ecor

\deleted{Note that the first JPL bound for $R(\delta)$ is
\[
R(\delta) \le H\(\frac12 - \sqrt{\delta(1-\delta)}\)
\]
Hence our bound constitutes an improvement.}

\noi We give \replaced{better}{explicit} estimates for $|B_{\T}(\rho n )|$ when $w = 3$ and $w = 4$, obtaining\deleted{, in particular,} the following \deleted{explicit}bounds for $R_3(\delta)$ \added{and $R_4(\delta)$}:

\thm
\label{thm:R3}
Let \deleted{$\rho = \frac 12 - \sqrt{\delta(1-\delta)}$, for} $0 \le \delta \le \frac12$. Then
\[
R_3(\delta) \le \left\{\begin{array}{lll}
\rho + \frac 12 H(2\rho)  & \mbox{if} & \delta \geq \frac 12 - \frac {\sqrt 2}{3} \\
\frac 13  + \frac12 H(1/3)   & & \mbox{otherwise}
\end{array}\right.
\]
\added{where $\rho = \frac 12 - \sqrt{\delta(1-\delta)}$}.
\ethm

\thm
\label{thm:R4}
\added{Let $0 \le \delta \le \frac12$. Then
\[ R_4(\delta) \le H(\rho) - \frac{\rho}{2} \cdot \log_2\(\frac{1}{(1-\rho)^4 + 4\rho(1-\rho)^3 + 6\rho^2(1-\rho)^2}\)\]
where $\rho = \frac 12 - \sqrt{\delta(1-\delta)}$}.
\ethm

\rem
\label{rem:3-4}
The bound for $R_3(\delta)$ looks different from the one predicted by Corollary~\ref{cor:LP-for-LDPC}. The reason is that we have a particular way to upper bound the metric balls in $\T$ for $w=3$, which provides better bounds than Corollary~\ref{cor:LP-for-LDPC}. \deleted{For a bound on $R_4(\delta)$ which behaves according to the corollary, see Corollary~\ref{cor:w=4} in the Appendix}
\erem
\deleted{Note that for $\delta < \frac 12 - \frac {\sqrt 2}{3} \approx 0.02586$ the upper bound $R_0 = \frac 13  + \frac12 H(1/3) \approx 0.7924$ does not depend on $\delta$. This is simply the maximal rate an LDPC code with $w = 3$ can have.}

\subsubsection*{Comparing with Known Bounds}
\replaced{Our} {We compare the} bound in Theorem~\ref{thm:R3} \replaced{is better than the best known} {with} bounds for $R_3(\delta)$ \deleted{in}\cite{ ben2006upper}, \replaced {when} {. Our bound is better for} $\delta$ \added{is} sufficiently close to $1/2$. However, we can do better. The argument in \cite{ben2006upper} holds if we replace the JPL bound \replaced{it uses}{(which that paper uses)} with our improved bound. This leads to a better bound on $R_3(\delta)$ for $0.156 < \delta < 0.5$.

\noi \deleted{Note that t}The same line of argument leads to improved bounds \added{on $R_w(\delta)$ for $0.287< \delta < 0.5$}, for any \replaced{$w > 3$}{$w \geq 3$}. \added{This range could probably be extended, but} \replaced {w}{W}e do not attempt \replaced{to do so}{ a detailed analysis of the interval of gain for $w >3$} in this paper.

\subsubsection*{Organization}
\deleted{This paper is organized as follows. }We prove Theorem~\ref{thm:tor_size} in Section~\ref{sec:metric balls}. \replaced{Theorems~\ref{thm:R3} and~\ref{thm:R4} are proved in Sections~\ref{sec:R3} and~\ref{sec:R4} respectively.} {Theorem~\ref{thm:R3} is proved in Section~\ref{sec:R3}.} Comparison with known bounds is done in Section~\ref{sec:compare}. \deleted{An explicit bound on $c(4,\rho)$ is given in the Appendix.}

\subsubsection*{\added{Notation}}
\added{Throughout the paper, given a vector $v \in \B$, we set $s(v)$ to be its support viewed as a subset of $\{1 \ldots n\}$.}

\section {Proof of Theorem~\ref{thm:tor_size}}
\label{sec:metric balls}

\deleted{We start by defining the {\it coset leader} for a coset $x + C^{\perp} \in \T$. This is an element of minimal weight in the coset. If the coset has more than one element of minimal weight, we take the coset leader to be minimal in the lexicographic order among these.}
\deleted{In particular, a coset leader is an element of $\B$. Note that the metric ball $B_{\T}(r)$ is the set of cosets with coset leader of Hamming weight at most $r$.}

\noi Our \replaced{first}{next} step reduces the problem to estimating a certain probability. Given $0 < \rho < 1/2$, let $x$ be a random vector in $\B$, obtained by setting the coordinates independently to $1$ with probability $\rho = r/n$ and to $0$ with probability $1-\rho$. Let $p = p(\rho)$ be the probability that $x$ is a coset leader. In the following discussion we may, and will, assume $\rho n$ is an integer.
\lem
\label{lem:prob}
\[
p(\rho) \ge \Omega\(\frac{1}{\sqrt{n}}\) \cdot \frac{|B_{\T}(\rho n)|}{|B(\rho n)|}
\]
\elem
\prf
Note that for $\rho < 1/2$ the function $f(k) = \rho^k (1-\rho)^{n-k}$ decreases in $k$. Recall also that (by Stirling's formula) $|B(\rho n)| \ge \Omega\(\frac{1}{\sqrt{n}}\) \cdot 2^{H(\rho) n}$. Therefore
\[
p(\rho) \ge |B_{\T}(\rho n)| \cdot \rho^{\rho n} (1-\rho)^{n-\rho n} = |B_{\T}(\rho n)| \cdot 2^{-H(\rho) n} \ge \Omega\(\frac{1}{\sqrt{n}}\) \cdot \frac{|B_{\T}(\rho n)|}{|B(\rho n)|}
\]
\eprf

\noi Hence, \replaced{the claim of the theorem reduces to showing} {we only need to argue} that there exists \added{an absolute constant} $c = c(w,\rho) \replaced{\geq \frac{\log_2e}{8w^2} \cdot \(\frac{\rho^w}{2}\)^{w+1}}{> 0}$ such that $p < 2^{-cn}$.

\noi Let $v_1,...,v_m$ be a basis of $C^{\perp}$ whose elements are vectors of Hamming weight at most $w$. \deleted{Identify each $v_i$ with its support, viewed as a subset of $\{1,...,n\}$, and} \replaced{A}{a}ssume, w.l.o.g, that $\cup_{i=1}^m ~\replaced{s(v_i)}{v_i} = \{1, \ldots, n\}$.

\noi We partition the coordinates $ \{1, \ldots, n\}$ into $w$ disjoint sets $I_w, I_{w-1}, \ldots I_1$\added{, in the following way}. Let $1 \leq k \leq w$. Suppose $I_w, I_{w-1} \ldots I_{k+1}$ are already defined, and let us define $I_k$.  Initialize $I_k = \emptyset$. Go over the vectors $v_i$. If $\replaced{s(v_i)}{v_i}$ has exactly $k$ coordinates outside $I_w \cup I_{w-1} \cup \ldots \cup I_{k+1} \cup I_k$, add them to $I_k$.

\noi \added {Note that $|I_k|$ is always a multiple of $k$ (in particular $|I_k|$ can be zero). For instance, if $C^\perp$ is spanned by the vectors $v_1 = \{1,1,1,0,0\}$, $v_2 = \{0,0,1,1,0\}$ and $v_3 = \{0,1,0,0,1\}$, then the partition is $I_3 = \{1,2,3\}$, $I_2 = \emptyset$ and $I_1 = \{4,5\}$.}

\lem \label{lem:large}
Let $A = \frac{2}{\rho^w}$. \deleted{(note $A \ge 4$) }There exists an index $1 \le k \le w$ such that
\[
|I_k| > \max\left\{A \cdot \sum_{j=k+1}^w |I_j|,~ \frac{n}{2w A^w}\right\}
\]
\elem
\prf
If not, \replaced{we will show}{it is easy to see by induction on $k$,} that \added{$|I_k| < \frac{n}{w}$} for all $1 \le k \le w$\replaced{,}{
\[
|I_{w-k+1}| \le \frac{n}{2w A^w} \cdot \sum_{j=0}^{k-1} A^j < \frac{n}{w},
\]}
contradicting the fact that $|I_1| + |I_2|  + \ldots + |I_w| = n$.

\noi \added{We note, for future reference, that $\rho <1/2$ implies $A > 2^{w+1}$. }

\noi \added{Let $S(k)$ stand for $A \cdot \sum_{j=k}^w |I_j|$. Note that our assumption is that for any $1 \leq k \leq w$ holds
\[|I_k| \leq \max\left\{S(k+1),~ \frac{n}{2w A^w}\right\}.\]
}

\noi \added {Since $S(1) = A\cdot n$ and $S(w+1) = 0$, there exists an index $1 \leq k_0 \leq w$ such that $S(k_0+1) \leq \frac{n}{2w A^w} < S(k_0)$.}

\noi \added {We consider two cases, $k \ge k_0$ and $k < k_0$.}

\begin{itemize}

\item \added{$k \geq k_0$:}

\noi \added{This is the easy case. We have $S(k+1) \leq \frac{n}{2w A^w}$, and hence $|I_k| \leq \frac{n}{2w A^w} < \frac{n}{w}$. We record for later use that, in particular, $|I_{k_0}| \leq \frac{n}{2w A^w}$.}

\item \added{$k < k_0$:}

\noi \added{We start with a few preliminary observations. First, in this case $\frac{n}{2w A^w} < S(k+1)$ and hence $|I_k| \leq S(k+1)$. This implies that $S(k) = A \cdot |I_k| + S(k+1) \leq (A+1) \cdot S(k+1)$.}

\noi \added{Next, we argue that $|I_k| \leq (A+1)^{k_0 - k -1} \cdot S\(k_0\)$. This follows from the observations above, by applying the inequality $S(m) \leq (A+1) \cdot S(m+1)$ repeatedly for $m = k+1,...,k_0 - 1$.}

\noi \added{To complete the proof we need two more simple facts. Recall, that the definition of $k_0$ gives $S(k_0 + 1) \le \frac{n}{2w A^w}$, and hence $S(k_0) = A \cdot |I_{k_0}| + S(k_0+1) \le (A+1) \cdot \frac{n}{2w A^w}$.}

\noi \added{Finally, note that since $A > 2^{w+1} > w \ge 3$, we have $(A+1)^{w-1} < e\cdot A^{w-1} <A^w$. Putting everything together gives
\[
|I_k| < (A+1)^{k_0 - k -1} \cdot S\(k_0\) \le (A+1)^{w-2} \cdot S\(k_0\) \le (A+1)^{w-1} \cdot \frac{n}{2w A^w} < \frac{n}{w}
\]}
\end{itemize}
\eprf

\noi Let $k$ be the index given by the lemma. Set $m = |I_k|$. Note that the coordinates of $I_k$ are divided into $t = m/k$ disjoint $k$-tuples \replaced{$U_1 \ldots U_t$}{$u_1 \ldots u_t$} and each \replaced{$U_i$}{$u_i$} is contained in \added{the support of} a \replaced{different} {distinct} basis element $v_{j_i}$. Note also that \replaced{$s(v_{j_i})\setminus U_i$}{$v_{j_i}\setminus u_i$} is a subset of $\cup_{j=k+1}^w I_j$.

\noi We claim that \added{the support of} any coset leader $x$ must contain at most $\frac{\rho^k}{2} \cdot t$ of the $k$-tuples \replaced{$U_i$}{$u_i$}. Indeed, assume not and let $S \subset \{1 \ldots t\}$ be the set of indices $i$ such that \replaced{$U_i \subset s(x)$}{$u_i \subseteq x$}. Let $y = x +\sum_{i \in S} v_{j_i}$.  Since \replaced{$s(x)$}{$x$} and \replaced{$s(y)$}{$y$} coincide on $I_1 \cup I_2 \cup \ldots \cup I_{k-1}$, we have
\[
|y| \le |x| - k \cdot |S| + \sum_{j=k+1}^w |I_j| < |x| - \frac{\rho^k}{2} \cdot |I_k| + \sum_{j=k+1}^w |I_j| < |x|
\]
where the last inequality follows from the choice of $k$. \replaced{Since} {On the other hand,} $y$ belongs to the same coset as $x$, \replaced {this contradicts} {contradicting} the fact that $x$ is a coset leader.

\noi Now, let $x$ be a random vector with coordinates set independently to $1$ with probability $\rho = r/n$ and to $0$ with probability $1-\rho$. Each \added{$k$-}tuple \replaced{$U_i$}{$u_i$} is in \replaced{$s(x)$}{$x$} with probability $\rho^k$ and the events of containing distinct tuples are statistically independent, since the tuples are disjoint. Let $p_0$ be the probability that \replaced{$s(x)$}{$x$} contains at most  $\frac{\rho^k}{2} \cdot t$ \replaced{of the tuples $U_1 \ldots U_t$}{tuples}. By the preceding discussion, \added{it upper bounds the probability $p$ that $x$ is a coset leader.} \replaced{Applying the Chernoff bound we have,}{and the Chernoff bound, we have, for the probability $p$ that $x$ is a coset leader,}
\[
p \le p_0 \le exp\left\{-\frac{\rho^k \cdot t}{8}\right\} \deleted{\le 2^{-c n}}
\]
\deleted{where $c = c(w,\rho)$ is a constant depending only on $\rho$ and $w$.} \replaced{Recall that $t = \frac{|I_k|}{k}$. From Lemma~\ref{lem:large}, $|I_k| > \frac{n}{2wA^w}$. Hence,
\[
\rho^k \cdot t = \frac{\rho^k}{k} \cdot \frac{|I_k|}{n}\cdot n \geq \frac{\rho^w}{w} \cdot\frac {1}{2wA^w}\cdot n = \frac{1}{w^2}\(\frac{\rho^w}{2}\)^{w+1} \cdot n
\]
\added{Hence $p \leq 2^{-cn}$ where} $c = c(w,\rho) \geq \frac{\log_2e}{8w^2} \cdot \(\frac{\rho^w}{2}\)^{w+1}$, completing the proof of the theorem.}{Taking a more detailed look at the estimates provided by Lemma~\ref{lem:large}, it is easy to see that $c = c(w,\rho) \geq \frac{\log_2e}{w^2} \cdot \(\frac{\rho^w}{2}\)^{w+1}$.}

\section{Proof of Theorem~\ref{thm:R3}}
\label{sec:R3}
In this section we \replaced{treat} { present a sui generis argument for} the case $w=3$. \replaced{We present a simple argument to bound the growth of metric balls in the coset leader graph $\T$, which does better in this special case than the more general approach of Theorem \ref{thm:tor_size}.} {This argument provides an explicit bound better than what we are able to derive following the line of argument in the proof of Theorem~\ref{thm:tor_size}.}\deleted{In the Appendix, we present a derivation for the case $w=4$, that follows the line of argument in the proof of Theorem~\ref{thm:tor_size}.} Unfortunately, we were not able to extend it to larger values of $w$.

\noi We will argue that for any distance $r$ attainable in $\T$, an element $x + C^{\perp}$ which belongs to the $r$-sphere $S_r = S_{\T}(r)$ around zero has at most $n-2r$ neighbours in the next sphere $S_{r+1}$. This should be compared to the \replaced{situation}{behavior} in the Hamming cube, in which an\deleted{y} element in the $r$-sphere has $n-r$ neighbours in the $(r+1)$-sphere. A simple calculation will then show that the metric balls in the coset leader graph grow \replaced{much}{exponentially} slower than in the cube, and prove the claim of the theorem.

\noi In the following discussion we \deleted{(again) identify a binary vector with its support, and}assume \replaced{w.l.o.g. that $\cup_{v\in C^\perp, |v| \leq 3}s(v) = \{1, \ldots, n\}$}{$\cup_{v \in C^\perp} {v} = \{1, \ldots, n\}$}.

\noi Consider an element $x + C^{\perp} \in S_r$\replaced{. Assume}{, where} $x$ is the coset leader, in particular $|x| = r$. For each coordinate $i \in \replaced{s(x)}{x}$ let $v_i \in C^\perp$ be a vector of weight at most $3$ \replaced{whose support contains $i$}{containing $i$}. The key point in the argument is that there are at least $2r$ directions to go from $x + C^{\perp}$  that {\it do not} lead away from zero. This is shown in the following \replaced{l}{L}emma.
\lem
\label{lem:3-other}
\begin{enumerate}
\item
For all $j \in \cup_{i \in \replaced{s(x)}{x}} \replaced{s(v_i)}{v_i} \quad \mbox{holds} \quad d\(0,x+e_j+C^{\perp}\) \le r$

\item $\Big |\cup_{i \in \replaced{s(x)}{x}} \replaced{s(v_i)}{v_i}\Big |\, \ge \, 2r \quad\mbox{ (in particular $r\leq n/2$)} $

\end{enumerate}
\elem
\prf
\deleted {First, note that $|{v_i \cap x}| = 1$, otherwise $y = x + v_i$ would be a smaller weight element in the same coset.}\added{First note that $s(x) \subseteq \cup_{i \in s(x)}s(v_i)$.} Let $j \in \cup_{i \in \replaced{s(x)}{x}} \replaced {s(v_i)}{v_i}$. \added{We distinguish between two cases.} If $j \in \replaced{s(x)}{x}$, the element $(x+e_j) + C^\perp$ is in $S_{r-1}$. For $j \not \in \replaced{s(x)}{x}$, let $j \in \replaced{s(v_i)}{v_i}$ for some $i\in \replaced{s(x)}{x}$. The vector $x+e_j + v_i$ is of weight at most $r$, since $i \in \replaced{s(x)}{x}$ \replaced{,}{and} $j \in \replaced {s(v_i)}{v_i}$\added{, and $i \not = j$}. Therefore $d\(0,x+e_j+C^{\perp}\) \le r$.

\noi It remains to show $\Big |\cup_{i \in \replaced{s(x)}{x}} \replaced{s(v_i)}{v_i}\Big | \ge 2r$.

\noi Let $z  = \sum_{i\in \replaced{s(x)}{x}} v_i$. \added{We will show that $|z| \geq 2r$, which will give what we want, since $z$ is supported in $\cup_{i \in s(x)} s(v_i)$}. \replaced{Observe that for all $i \in s(x)$ holds}{Since} $\replaced{s(v_i) \cap s(x)}{v_i \cap x} =\{i\} $, \added{since otherwise $y = x + v_i$ would be a smaller weight element in the same coset.} \replaced{Hence $s(x) \subseteq s(z)$, which}{clearly $x \subseteq z$. Note that this} implies $|z| \geq 2r$. \replaced{Indeed, if not}{Otherwise}, we would have $|x+z| = |z| - |x| < r$, and $y = x+z$ would be a smaller weight element in the coset \added {of $x$}. \deleted {Since $z$ is supported in $|\cup_{i \in x} v_i|$, we have $|\cup_{i \in x} v_i| \ge 2r$, as required.}
\eprf

\noi We now use this to bound the rate of growth of metric spheres in $\T$. Consider the bipartite graph whose parts are given by $S_r$ and $S_{r+1}$ and two vertices are connected if they are neighbours in $\T$. We have shown that the degree of any element in $S_r$ is at most $n-2r$. On the other hand, the degree of every element in $S_{r+1}$ is, obviously, at least $r+1$. By a standard double counting argument, this implies
\[
|S_{r+1}| \leq \frac{n-2r}{r+1} \cdot |S_r|
\]
Therefore, for $r \le n/2$ holds
\[
|S_r| \leq \frac{1}{r!} \cdot \prod_{k=0}^{r-1} (n-2k) \le 2^r \cdot \binom {\lceil n/2 \rceil}{r}
\]
and, obviously, $S_r = 0$ for larger $r$.

\noi The expression $2^r \cdot \binom {\lceil n/2 \rceil}{r}$ increases in $r$ till $r = \lceil n/3 \rceil$ and decreases for larger $r$. Therefore (omitting\deleted{the} integer rounding for \added{the sake of} typographic clarity)

\[
|B_{\T}(r)| ~~=~~ \sum_{k=0}^r |S_k| \quad < \quad \left\{
                                       \begin{array}{ll}
                                         n \cdot 2^r  \cdot \binom {n/2}{r} & \hbox{if $r\leq  n/3 $;} \\
                                         n \cdot 2^{ n/3}  \cdot \binom { n/2 }{ n/3 }  & \hbox{if $r >  n/3 $.}
                                       \end{array}
                                     \right.  \]

\noi Substituting $r = \rho n$ and using the inequality $\binom {n} {\rho n} \le 2^{n H(\rho)}$, we obtain
\beqn
\label{second-prf}
|B_{\T}(\rho n)| \leq \left\{
\begin{array}{ll}
2^{n\(\rho + \frac12 H(2\rho) \)}      & \hbox{if $\rho \leq 1/3$;} \\
2^{n\(\frac 13  + \frac12 H(2/3)\) } & \hbox{if $\rho > 1/3$.}
\end{array}
\right.
\eeqn

\noi \replaced{The proof of Theorem~\ref{thm:R3} is completed by using this bound in Theorem~\ref{thm:FT}.} {This completes the proof of Theorem~\ref{thm:R3}}
\eprf

\section{\replaced{Proof of Theorem~\ref{thm:R4}}{Bounds on $|B_{\T}(r)|$ for $w=4$} }
\label{sec:R4}

\noi \replaced{We deduce the theorem from Theorem~\ref{thm:FT} by showing that $|B_{\T}(\rho n)| \le 2^{-cn} \cdot |B(\rho n)|$ where}{The main result of this section is an explicit lower bound on $c(4,\rho)$, the constant in Theorem~\ref{thm:tor_size}. We show}
\beqn
\label{eq:c4}
c = c(\rho) \ge \frac{\rho}{2} \cdot \log_2\(\frac{1}{(1-\rho)^4 + 4\rho(1-\rho)^3 + 6\rho^2(1-\rho)^2}\) - o_n(1)
\eeqn
\deleted{This implies the corresponding upper bound on $R_4(\delta)$ via Corollary~\ref{cor:LP-for-LDPC}.}

\deleted{Let $\rho = \sqrt{\frac12 - \delta(1-\delta)}$. Then
\[
R_4(\delta) \le H(\rho) - \frac{\rho}{2} \cdot \log_2\(\frac{1}{(1-\rho)^4 + 4\rho(1-\rho)^3 + 6\rho^2(1-\rho)^2}\)
\]}

\noi Let $I_1, I_2, I_3, I_4$ be the partition of $[n]$ \added{, as} defined in the proof of Theorem~\ref{thm:tor_size}. \replaced{Let}{Denote:}
\[
|I_1| = \alpha_1 n,\quad |I_2| = \alpha_2 n, \quad |I_3| = \alpha_3 n, \quad |I_4| = \alpha_4 n
\]
Note that $\alpha_1 + \alpha_2 + \alpha_3 + \alpha_4 = 1$.

\noi \replaced{The following lemma shows existence of}{We start with a lemma that shows that we can find} elements of\deleted{a} prescribed structure in each coset of $C^{\perp}$. Both\deleted{in} the statement and\deleted{in} the proof of the lemma\deleted{we} refer to the properties of \replaced{the partition $\{I_j\}$, }{$I_j$} as described in the proof of Theorem~\ref{thm:tor_size}.

\lem
\label{lem:struct}
Let $u \in \B$.
\begin{enumerate}
\item
There is an element $u_1 \in u + C^{\perp}$ \replaced{whose support}{that} does not intersect $I_1$.
\item
There is an element $u_2 \in u + C^{\perp}$ whose weight is at most that of $u$ and such that
\begin{itemize}
\item $\replaced{s(u_2)}{u_2} \cap I_1 = \replaced{s(u)}{u} \cap I_1$.
\item $\replaced{s(u_2)}{u_2}$ intersects each \replaced{$j$-tuple}{pair} of \replaced{$I_j$}{$I_2$} in a most \replaced{$\lfloor j/2 \rfloor$}{one} coordinate\added{s for $j = 2,3,4$}.
 \deleted{$u_2$ intersects each triple of $I_3$ in at most one coordinate.}
 \deleted{$u_2$ intersects each $4$-tuple of $I_4$ in at most two coordinates.}
\end{itemize}
\end{enumerate}
\elem

\noi Before proving the lemma, we state two corollaries.
\cor
\label{cor:w=4}
\begin{enumerate}
\item Each coset of $C^\perp$ has a representative \replaced{whose support intersects each $j$-tuple of $I_j$ in a most $\lfloor j/2 \rfloor$ coordinates for $j = 1,2,3,4$.}{that does not intersect $I_1$ at all; intersects each pair of $I_2$ in a most one coordinate; intersects each triple of $I_3$ in at most one coordinate; and intersects each $4$-tuple of $I_4$ in at most two coordinates.}

\item Each coset of $C^\perp$ has a minimal weight representative \replaced{whose support intersects each $j$-tuple of $I_j$ in a most $\lfloor j/2 \rfloor$ coordinates for $j = 2,3,4$.}{that intersects each pair of $I_2$ in a most one coordinate; intersects each triple of $I_3$ in at most one coordinate; and intersects each $4$-tuple of $I_4$ in at most two coordinates.}
\end{enumerate}
\ecor
\prf
\begin{enumerate}
\item Apply both parts of the lemma to any element $u$ in the coset.
\item Apply the second part of the lemma to a minimal weight element $u$ in the coset.
\end{enumerate}
\eprf

\cor
\label{cor:diam}
The diameter of the coset leader graph $\T = \B/C^{\perp}$ is at most $D = \(\alpha_2/2 + \alpha_3/3 + \alpha_4/2\) \cdot n$.
\ecor
\prf

\noi Since $\T$ is vertex-transitive, it suffices to show that the distance of any coset of $C^{\perp}$ from zero is at most $D$. \replaced{To see this, note that each coset has a representative whose}{That is, each coset of $C^{\perp}$ contains an element of weight at most $D$.Indeed, any coset has a representative with a} structure \added{is} given by the first part of Corollary~\ref{cor:w=4}. It is immediate that its weight is at most $D$.
\eprf

\noi {\bf Proof of Lemma~\ref{lem:struct}}

\noi The first part of the lemma. For each coordinate $i \in I_1$ contained in \added{the support of} $u$, add to $u$ a basis vector $v_i \in C^\perp$ \replaced{whose support}{that} intersects $I_1$ only in this coordinate. \added{Such a vector exists from the definition of $I_1$.} This process \replaced{terminates}{results} in an element $u_1$ in the same coset, whose \replaced{support does not intersect $I_1$.} {intersection with $I_1$ is empty.}

\noi The second part of the lemma. We modify $u$ in three steps, by adding vectors from $C^{\perp}$, until we arrive to the \replaced{required}{desired} structure. We \replaced{keep track of}{observe that} the weight of $u$ \added{to see that it} does not increase in the process.
\begin{enumerate}
\item
For each pair $(i,j)$ in $I_2$ contained in \added{the support of} $u$, add to $u$ a basis vector $v \in C^\perp$ of weight at most four \replaced{whose support contains $(i,j)$}{containing the pair}, \replaced {and whose remaining elements} {whose remaining $1$-coordinates} are in $I_3 \cup I_4$. Note that this does not increase the weight of $u$ and does not \replaced{change}{affect} its intersection with $I_1$.
At the end of this step we \replaced{obtain}{arrive to} an element $u' \in u + C^{\perp}$ \replaced{whose support intersects} {intersecting} each pair of $I_2$ in at most one coordinate.

\item
For each triple in $I_3$ that intersects \added{the support of }$u'$ in at least two coordinates, add to $u'$ a basis vector $v\in C^\perp$ of weight at most four \replaced{whose support}{that} contains \replaced{this}{the} triple, and whose remaining \replaced{element}{$1$-coordinate} (if it exists) is in $I_4$. \replaced{This}{Note that this, again,} does not increase the weight of $u'$ and does not \replaced{change}{affect} its intersection with $I_1$ and $I_2$.
This step terminates at an element $u''$ of the same coset \replaced{intersecting}{whose intersections with} $I_1$, $I_2$ and $I_3$\deleted{are} as required.

\item
For each $4$-tuple in $I_4$ that intersects \added{the support of} $u''$ in more than two coordinates, add it to $u''$.\deleted{As above,} This does not increase the weight of $u''$ and does not \replaced{change}{affect} its intersection with $I_1$, $I_2$, \added {and} $I_3$. At the end of the process we obtain an element $u_2 \in u + C^{\perp}$ \replaced{intersecting all $I_j$} {whose intersections with $I_j$, $j = 1,\ldots,4$ are} as required.
\end{enumerate}

\eprf

\noi We proceed \replaced{towards the proof of (\ref{eq:c4}).}{to upper bound the cardinalities $|B_{\T}(r)|$ of metric balls in $\T$.} By Corollary~\ref{cor:diam}, it suffices to \replaced{deal with $0 \le \rho\ \le \frac {\alpha_2}{2} + \frac {\alpha_3}{3} + \frac{\alpha_4}{2}$.}{analyze $|B_{\T}(r)|$ for $r \leq \(\frac {\alpha_2}{2} + \frac {\alpha_3}{3} + \frac{\alpha_4}{2}\) \cdot n$.} Let us fix such \replaced{$\rho$}{$r$ and set $\rho = r/n$}.

\noi \replaced{Let $x$ be a random vector in $\B$, obtained by setting the coordinates independently to $1$ with probability $\rho = r/n$ and to $0$ with probability $1-\rho$. Let $p = p(\rho)$ be the probability that $x$ is a coset leader. By Lemma~\ref{lem:prob} it is enough to show $p(\rho) \leq 2^{-cn}$ where  $c$ is given by the RHS of (\ref{eq:c4}).}{Consider the following probabilistic experiment. Let $x \in \B$ be a random vector whose coordinates are independently set to $1$ with probability $\rho$ and to $0$ with probability $1-\rho$. Let $p = p(\rho)$ be the probability that $x$ is of minimal weight in its coset and has the structure described in the second part of Corollary~\ref{cor:w=4}. Note that each coset has exactly one coset leader and at least one element with the properties described in the corollary. Therefore $p$ is greater or equal than the probability of $x$ to be a coset leader. Hence, by Lemma~\ref{lem:prob}, it is enough to upper bound $p$.}

\noi \added{Let $p'= p'(\rho)$ be the probability that $x$ is of minimal weight in its coset and has the structure prescribed by the second part of Corollary~\ref{cor:w=4}. Note that each coset has exactly one coset leader and at least one element with the properties given in the corollary. Therefore $p \leq p'$. In the remaining part of the proof we show that $p' \leq 2^{-cn}$.}

\noi \replaced{Corollary~\ref{cor:w=4} imposes $\(\frac {\alpha_2}{2} + \frac {\alpha_3}{3} + \frac{\alpha_4}{4}\) \cdot n$ statistically independent}{The} constraints on $x$\deleted{ arising from its constrained intersections with elements of $I_j$, $j > 1$, are independent}. The probability for all of them to hold is
\beqn
\label{eq:prob}
\((1-\rho)^4 + 4\rho(1-\rho)^3 + 6\rho^2(1-\rho)^2\)^{\frac 14 \alpha_4 n} \cdot \((1-\rho)^3 + 3\rho(1-\rho)^2\)^{\frac 13 \alpha_3 n}\cdot \(1-\rho^2\)^ {\frac 12 \alpha_2 n}
\eeqn

\noi \replaced{Recall that $\rho \leq \alpha_2/2 + \alpha_3/3 +\alpha_4/2$. Hence, $p'(\rho)$ is bounded from above by the maximum value the expression in (\ref{eq:prob}) attains in the domain}{We want to maximize this expression over the domain}
\[
\Delta(\rho) = \Big\{\alpha_1,  \alpha_2, \alpha_3, \alpha_4 \geq 0,\quad \alpha_1 + \alpha_2 + \alpha_3 + \alpha_4 = 1,\quad \alpha_2/2 + \alpha_3/3 +\alpha_4/2 \ge \rho \Big\}
\]

\noi We \replaced{claim}{argue} that for fixed $\alpha_1$, this expression is maximized when $\alpha_2 = \alpha_3 =0$ and $\alpha_4 = 1 - \alpha_1$. \replaced{To see this, w}{W}e start with a technical lemma:

\lem
For any $0 \le \rho \le 1/2$,
\[
\((1-\rho)^4 + 4\rho(1-\rho)^3 + 6\rho^2(1-\rho)^2\)^{\frac 14} \ge \max\left\{\((1-\rho)^3 + 3\rho(1-\rho)^2\)^{\frac 13}, \(1-\rho^2\)^ {\frac 12} \right\}
\]
\elem

\prf
Dividing out by $(1-\rho)^{1/2}$ and rearranging, it suffices to show:
\[ \((1+\rho)^2 + 2 \rho^2 \)^{1/4}\geq \(1+\rho\)^{1/2} \geq (1-\rho)^{1/6}(1+2\rho)^{1/3}\]

\noi The first inequality \replaced{is immediate.}{follows by comparing the fourth power of the two terms.} For the second inequality, observe that
\[(1+\rho)^3 - (1-\rho)(1+2\rho)^2 = \replaced{5\rho^3 + 3\rho^2}{\rho^2 - \rho^3} \geq 0\]
\eprf

\noi By the lemma, increasing $\alpha_4$ and decreasing $\alpha_2 + \alpha_3$ by the same amount increases (\ref{eq:prob}) and leaves us in $\Delta(\rho)$ as long as $\alpha_2,\alpha_3\ge 0$. \replaced{Consequently}{Hence}, we may take $\alpha_2 = \alpha_3 = 0$ and $\alpha_4 = 1 - \alpha_1$.

\noi We arrive to the problem of maximizing $\((1-\rho)^4 + 4\rho(1-\rho)^3 + 6\rho^2(1-\rho)^2\)^{\frac 14 \alpha_4 n}$ on $[2\rho,1]$.

\noi Since $(1-\rho)^4 + 4\rho(1-\rho)^3 + 6\rho^2(1-\rho)^2 < 1$, the maximum is attained at $\alpha_4 = 2\rho$. Hence
\[
p\added{'} \leq \((1-\rho)^4 + 4\rho(1-\rho)^3 + 6\rho^2(1-\rho)^2\)^{\frac 12 \rho n},
\]
concluding the proof of (\ref{eq:c4}).

\section{Comparison to Other Bounds \deleted{for $w=3$} } \label{sec:compare}

\noi Ben Haim and Litsyn \cite{ben2006upper}, (see also \cite{burshtein2002upper}) give the best known upper bounds on the rate of LDPC codes \added{with relative minimal distance $\delta$}:\footnote{$R_{LP}$ is the second JPL bound. For the definition of $R_{cw}$ see \cite{ben2006upper}.}
\begin{eqnarray}
  R(C) &\leq R\added{_w}^{(1)}(\delta)& = 1 - \frac {H(\delta/2)}{H((1- (1-\delta)^w)/2)}\label{eq:T1}\\
  R(C) &\leq R\added{_w}^{(2)}(\delta)& = 1 - \max_{\delta/2 \leq u \leq 1/2} \(\frac{H(u) - R_{cw}(u,\delta)}{H((1-(1-2u)^w)/2)}\)  \label{eq:T2}\\
  R(C) &\leq R\added{_w}^{(4)}(\delta)& =\min_{0 \leq t \leq 1-2\delta} \((1-t)R_{LP}(\delta / (1-t)) +t - \frac{t}{w}\)  \label{eq:T4} \\
  R(C) &\leq R\added{_w}^{(5)}(\delta)& =\min_{0 \leq t \leq 1-2\delta} \((1-t)R_{LP}(\delta / (1-t)) +t - \frac{t}{w-1}\) \label{eq:T5}
\end{eqnarray}
where $R\added{_w}^{(1)}(\delta)$, $R\added{_w}^{(2)}(\delta)$, $R\added{_w}^{(4)}(\delta)$, and $R\added{_w}^{(5)}(\delta)$ are the bounds in Theorems 1, 2, 4 and 5 respectively of \cite{ben2006upper}. \added{Let us mention that the bound $R_w^{(5)}(\delta)$ requires an additional assumption, namely that the weight of each column in the parity check matrix is at least two.}

\subsection{\added{The Case $w=3$}}

\begin{figure}[!ht]
\centering
\includegraphics[width=114mm, natwidth=815, natheight= 615]{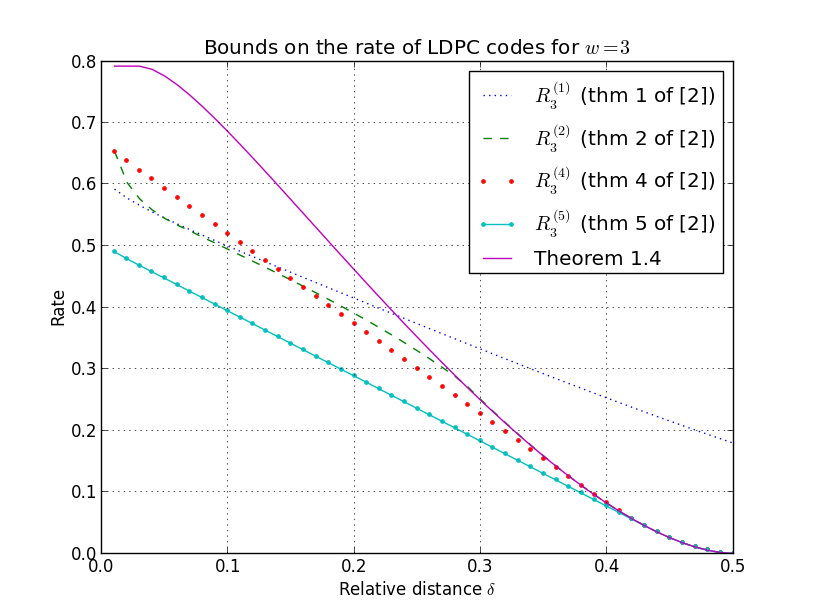}
\caption{The bound in Theorem~\ref{thm:R3} and the bounds of \cite{ben2006upper}  }
\label{fg:ours_vs_LB}
\end{figure}

\noi \replaced{Figure~\ref{fg:ours_vs_LB} presents several}{The next image visualizes the different} bounds for the case $w=3$.

\noi \replaced{We start with a comparison between the different bounds from \cite{ben2006upper}. The bound $R_3^{(5)}$ is better than the others}{The combined plots above require an explanation. First, $R_3^{(5)}$ is the best of \cite{ben2006upper} bounds} for the whole range \replaced{$0 \leq \delta \leq 0.5$}{$0 < \delta <0.5$}. However, it requires \replaced{the}{an} additional assumption\deleted{, namely} that the weight of each column in the parity matrix is at least 2. Without this assumption, we are left with the bounds $R\added{_3}^{(1)}$, $R\added{_3}^{(2)}$ and $R\added{_3}^{(4)}$, each of which is optimal in a subrange of \replaced{$0 \leq \delta \leq 0.5$}{$0 < \delta <0.5$}.

\noi Our bound \added{in Theorem~\ref{thm:R3}} is better than $R\added{_3}^{(4)}$ for $\delta > 0.3877$ and better than $R\added{_3}^{(5)}$ for $\delta > 0.4387$, since for these values of $\delta$ those two bounds \replaced{coincide with}{are very close to} the first JPL bound, and the bound in Theorem~\ref{thm:R3} is always better than the first JPL bound.

\noi \replaced{With that}{However}, we can do better. The argument in Theorems 4 and 5 in \cite{ben2006upper} holds if we replace \replaced{the second JPL bound they use}{$R_{LP}$} with the better bound of Theorem~\ref{thm:R3} (since the first and the second JPL bounds coincide \replaced{at the optimal values of $t$ in (\ref{eq:T4}) and (\ref{eq:T5})}{in this case}). This leads to a (small\footnote{\added{of magnitude $10^{-4}$ - $10^{-5}$}}) improvement on $R\added{_3}^{(4)}$ and $R\added{_3}^{(5)}$, and hence to best known bounds when these two bounds are optimal. ($R\added{_3}^{(4)}$ is optimal for $0.156 < \delta < 1/2$).

\noi To sum up, we improve the bounds on $R_3(\delta)$ for $0.156 < \delta < 1/2$. Given the additional assumption that the weight of each column in the parity check matrix is at least 2, we improve the bounds on the rate for the whole range $0 < \delta <0.5$.

\subsection{\added{The Case $w>3$}}

\noi \added{In this subsection, for brevity's sake, we deal only with bounds on $R_w(\delta)$, with no additional assumptions on the weight of the columns in the parity check matrix. Consider the subrange of the interval $0 <\delta < 0.5$ in which the following two conditions hold. The bound $R_w^{(4)}(\delta)$ of \cite{ben2006upper} is better than $R_w^{(1)}(\delta)$ and $R_w^{(2)}(\delta)$, and in addition to this, the first and the second JPL bounds coincide at the optimal values of $t$ in (\ref{eq:T4}). In this subrange, similarly to the case $w=3$, we can use Corollary~\ref{cor:LP-for-LDPC} or Theorem \ref{thm:R4} to improve on $R_w^{(4)}(\delta)$, and hence on $R_w(\delta)$.}

\noi \added{We proceed by comparing the three bounds from \cite{ben2006upper}. For this purpose, we first compare them to the second JPL bound:  }

\begin{itemize}
  \item \added{$R_w^{(1)}(\delta)$:}

  \noi \added{Numerical calculations show that $R_3^{(1)}(\delta)$ is bigger than the second JPL bound for $\delta > 0.23$. Since $R_w^{(1)}(\delta)$ increases with $w$, this holds for all $w \geq 3$.}

  \item \added{$R_w^{(2)}(\delta)$:}

  \noi \added{Numerical calculations show that $R_3^{(2)}(\delta)$ equals to the second JPL bound for $\delta > 0.287$. Since $R_w^{(2)}(\delta)$ increases with $w$, it is at least as large as the second JPL bound for all $w \geq 3$ in this range.}

  \item \added{$R_w^{(4)}(\delta)$:}

  \noi \added{Substituting $t=0$ in the RHS of (\ref{eq:T4}) recovers the second JPL bound. Hence $R_w^{(4)}(\delta)$ is at most as large as the second JPL bound for all $w \geq 3$.}

\end{itemize}

\noi \added{To sum up: For $\delta > 0.287$, $R_w^{(4)}(\delta)$ is at least as good as the other two bounds, for all $w \geq 3$. Next, note that in this range, the first and the second JPL bounds coincide at all values of $t$ in (\ref{eq:T4}), since they coincide on the interval $0.273 \leq \delta \leq 0.5$. Hence, we improve the bounds of \cite{ben2006upper} in this range.}

\section*{Acknowledgment}

\added{We would like to thank the anonymous referees for their numerous suggestions that led to a significant improvement in the presentation of this paper.}

\bibliographystyle{ieeetr}
\bibliography{eran}

\begin{thebibliography}{1}

\bibitem{friedman2006generalized}
J.~Friedman and J.-P. Tillich, ``Generalized {A}lon-{B}oppana theorems and
  error-correcting codes,'' {\em SIAM J. Discrete Math.}, vol.~19, no.~3,
  pp.~700--718 (electronic), 2005.

\bibitem{ben2006upper}
Y.~Ben-Haim and S.~Litsyn, ``Upper bounds on the rate of {LDPC} codes as a
  function of minimum distance,'' {\em IEEE Trans. Inform. Theory}, vol.~52,
  no.~5, pp.~2092--2100, 2006.

\bibitem{macwilliams1977theory}
F.~J. MacWilliams and N.~J.~A. Sloane, {\em The theory of error-correcting
  codes}.
\newblock North-Holland Publishing Co., Amsterdam-New York-Oxford, 1977.
\newblock North-Holland Mathematical Library, Vol. 16.

\bibitem{mceliece1977new}
R.~J. McEliece, E.~R. Rodemich, H.~Rumsey, Jr., and L.~R. Welch, ``New upper
  bounds on the rate of a code via the {D}elsarte-{M}ac{W}illiams
  inequalities,'' {\em IEEE Trans. Information Theory}, vol.~IT-23, no.~2,
  pp.~157--166, 1977.

\bibitem{delsarte1973algebraic}
P.~Delsarte, ``An algebraic approach to the association schemes of coding
  theory,'' {\em Philips Res. Rep. Suppl.}, no.~10, pp.~vi+97, 1973.

\bibitem{gallager1963low}
R.~Gallager, ``Low-density parity-check codes,'' {\em MIT press}, 1963.

\bibitem{burshtein2002upper}
D.~Burshtein, M.~Krivelevich, S.~Litsyn, and G.~Miller, ``Upper bounds on the
  rate of {LDPC} codes,'' {\em IEEE Trans. Inform. Theory}, vol.~48, no.~9,
  pp.~2437--2449, 2002.

\end{thebibliography}

\end{document}